\documentclass[letter]{aa} 
\usepackage{graphicx}
\usepackage{txfonts}
\begin{document}

\title{ The distribution of water in the high-mass star-forming region
  NGC~6334~I \thanks{\emph{Herschel} is an ESA space observatory with science
    instruments provided by European-led principal Investigator
    consortia and with important participation from NASA}}

   \author{M.~Emprechtinger
          \inst{1,20}
          \and
          D.~C.~Lis\inst{1}
	  \and 
T.~Bell\inst{1} \and 
T.~G.~Phillips\inst{1} \and
P.~Schilke\inst{10,20} \and
C.~Comito\inst{10} \and
R.~Rolffs \inst{10} \and
F.~van~der~Tak\inst{16} \and 
C.~Ceccarelli\inst{3} \and
H.~Aarts\inst{16}
A.~Bacmann\inst{3,2} \and
A.~Baudry\inst{2} \and
M.~Benedettini\inst{4} \and
E.A.~Bergin\inst{25} \and
G.~Blake\inst{1} \and
A.~Boogert\inst{5} \and
S.~Bottinelli\inst{6} \and
S.~Cabrit\inst{8} \and
P.~Caselli\inst{7} \and
A.~Castets\inst{3} \and
E.~Caux\inst{6} \and
J.~Cernicharo\inst{9} \and
C.~Codella\inst{11} \and
A.~Coutens\inst{6} \and
N.~Crimier\inst{3,9} \and
K.~Demyk\inst{6} \and
C.~Dominik\inst{12,13} \and
P.~Encrenaz\inst{8} \and
E.~Falgarone\inst{8} \and
A.~Fuente\inst{14} \and
M.~Gerin\inst{8} \and
P.~Goldsmith\inst{15} \and
F.~Helmich\inst{16} \and
P.~Hennebelle\inst{8} \and
T.~Henning\inst{26} \and
E.~Herbst\inst{17} \and
P.~Hily-Blant\inst{3} \and
T.~Jacq\inst{2} \and
C.~Kahane\inst{3} \and
M.~Kama\inst{12} \and
A.~Klotz\inst{6} \and
J.~Kooi\inst{1}
W.~Langer\inst{15} \and
B.~Lefloch\inst{3} \and
A.~Loose\inst{27}
S.~Lord\inst{5} \and
A.~Lorenzani\inst{11} \and
S.~Maret\inst{3} \and
G.~Melnick\inst{18} \and
D.~Neufeld\inst{19} \and
B.~Nisini\inst{24} \and
V.~Ossenkopf\inst{20}
S.~Pacheco\inst{3} \and
L.~Pagani\inst{8} \and
B.~Parise\inst{10} \and
J.~Pearson\inst{15} \and
C.~Risacher\inst{16}
M.~Salez\inst{8} \and
P.~Saraceno\inst{4} \and
K.~Schuster\inst{21} \and
J. Stutzki\inst{20}
X.~Tielens\inst{22} \and
M.~van der Wiel\inst{16} \and
C.~Vastel\inst{6} \and
S.~Viti\inst{23} \and
V.~Wakelam\inst{2} \and
A.~Walters\inst{6} \and
F.~Wyrowski\inst{10} \and
H.~Yorke\inst{15} 
          }

   \date{Received May 31, 2010; accepted July 23, 2010}

 
  \abstract
   {}
   {We present observations of twelve rotational transitions of
     H$_2^{16}$O, H$_2^{18}$O, and H$_2^{17}$O toward the massive
     star-forming region NGC~6334~I, carried out with \emph{Herschel/HIFI} as
     part of the guaranteed time key program \emph{Chemical HErschel
       Surveys of Star forming regions (CHESS)}. We analyze these
     observations to obtain insights into physical processes in this
     region.}
   {We identify three main gas components (hot core, cold foreground,
     and outflow) in NGC~6334~I and derive the physical conditions in
     these components. }
   {The hot core, identified by the emission in highly excited lines,
     shows a high excitation temperature of $\sim200$~K, whereas water
     in the foreground component is predominantly in the ortho- and
     para- ground states.
     The abundance of water varies between $4\cdot10^{-5}$ (outflow)
     and $10^{-8}$ (cold foreground gas). This variation is most
     likely due to the freeze-out of water molecules onto dust grains.
     The H$_2^{18}$O/H$_2^{17}$O abundance ratio is 3.2, which is consistent
     with the $^{18}$O/$^{17}$O ratio determined from CO
     isotopologues. The ortho/para ratio in water appears to be
     relatively low ($1.6\pm1)$ in the cold, quiescent gas, but close
     to the equilibrium value of three in the warmer outflow material
     ($2.5\pm0.8$).}
   {}
\institute{(Affiliations can be found after the references)}

   \maketitle
%

\section{Introduction}

Water is one of the most important coolants in star-forming regions,
and is thus a key molecule in the process of star formation. Water
plays a crucial role in the energy balance in protostellar objects, and
therefore its abundance is an important parameter, especially in
massive star-formation, because it influences the collapse and
fragmentation of cloud cores. Furthermore H$_2$O (either in the gas
phase or as ice on dust grains) is thought to contain a significant
fraction of oxygen. Thus well constrained water abundances provide
key input for astrochemical models of star-forming regions.

Most rotational lines of H$_2$O, including the ground-state transition
of ortho- and para-H$_2$O, are not observable from the ground.
However, previous space missions have already provided a glimpse of the
water universe (ISO, van Dishoeck \& Helmich~\cite{vH96}; SWAS,
Melnick \& Bergin~\cite{MB05}; Odin, Bjerkeli et al.~\cite{BLO09}).
The derived water abundance varies widely, between $\sim10^{-4}$ in
warm ($>100$~K) gas and $\sim10^{-8}$ in cold regions (e.g., Boonman
et al.~\cite{BDv03}). These differences in the H$_2$O abundance are
caused by the freeze-out of water molecules onto dust grains.
Unfortunately, all the observations to date suffer from a shortage of
accessible water lines (SWAS and Odin observed the ortho ground-state
transition only, while ISO only detected excited lines) and low
spatial resolution ($> 1'$).

Van der Tak et al.~(\cite{vdtak2010}) observed the para ground-state
transition ($1_{11}-0_{00}$) of H$_2^{16}$O in the high-mass
protostellar object DR~21(Main) with the \emph{Herschel} Space
Observatory (Pilbratt et al.~\cite{P10}). They found water abundances
between $10^{-10}$ and $10^{-7}$, about a factor 1000 lower than those
derived in previous studies. This shows that the H$_2$O abundance in
star-forming regions is not well understood and that studies including
multiple lines of both ortho- and para-H$_2$O are highly desirable to
put better constraints on the water abundance in the ISM. These
capabilities are provided by the HIFI instrument (de Graauw et
al.~\cite{dG10}).

In this letter, we present high spectral resolution observations of
twelve H$_2^{16}$O, H$_2^{17}$O, and H$_2^{18}$O lines in the
high-mass star-forming region NGC~6334~I. The relatively
nearby (1.7 kpc; Neckel~\cite{N78}) massive star-forming region NGC~6334
harbors sites of various stages of protostellar evolution (Straw \&
Hyland~\cite{SH89}). Single-dish continuum observations at
submillimeter wavelength revealed a total mass of 16700~M$_{\odot}$
(Matthews et al.~\cite{MMW08}), of which 200~M$_{\odot}$ is associated
with NGC~6334~I (Sandell~\cite{S00}). NGC~6334~I, studied extensively
over the last decades (e.g., Beuther et al.~\cite{BWT07, BWT08};
Hunter et al.~\cite{HBM06}), is a molecular hot core associated with a
cometary-shaped ultra compact H\,{\sc ii} (UCHII) region (de Pree et
al.~\cite{dPD95}), which exhibits a spectrum with very many emission lines
(Schilke et al.~\cite{SCT06}; Thorwirth et al.~\cite{TWM03}).
Furthermore, a bipolar outflow (Leurini et al.~\cite{LSP06}; Beuther
et al.~\cite{BWT08}) and several masers have been detected (e.g.,
Kreamer \& Jackson~\cite{KJ95}; Ellingsen et al.~\cite{EvM96}, Walsh
et al~\cite{WLT07}). Observations using the SMA showed that the hot
core of NGC~6334~I itself consists of four compact condensations
located within a $\approx 10''$ diameter region, which are emitting
about 50\% of the continuum detected in single-dish observations
(Hunter et al.~\cite{HBM06}).
 


\section{Observations}

We observed twelve water lines with frequencies between 488~GHz and
1.113~THz, which are listed in Table~\ref{lines}. These observations
were conducted between March 1 and March 23, 2010, using
\emph{Herschel}/HIFI in the dual beam switch (DBS) mode as part of the
guaranteed time key program {\it{Chemical HErschel Spectral Surveys
    (CHESS)}}. The coordinates of the observed position in NGC~6334~I
are $\alpha_{2000}: 17^h20^m53.32^s$ and $\delta_{2000}: -35^\circ
46'58.5''$. We used the Wide Band Spectrometer (WBS) with a resolution
of 1.1~MHz across a 4 GHz IF bandwidth. The spectra shown here are
equally weighted averages of the H and V polarization, reduced with the
HIPE pipeline (Ott~\cite{O10}), version 2.6. We exported the resulting
Level 2 double side band (DSB) spectra to the FITS format for a
subsequent data reduction and analysis with the IRAM GILDAS package.

The spectral scans of NGC~6334~I consist of DSB spectra with a
redundancy of eight, i.e., the lower and upper side band are observed with
eight different local oscillator settings. This observing mode allows the
deconvolution and isolation of the single sideband (SSB) spectra
(Comito \& Schilke~\cite{CS02}), which we present here.

The HIFI beam size at the frequencies observed is given in
Table~\ref{lines}. For the main beam efficiency we assumed a value of
68\%. The velocity calibration in HIFI data processed with HIPE~2.6 is
subject to uncertainties of up to $\sim 2$~km\ s$^{-1}$ due to an
approximate model for the spacecraft velocity used. A comparison
    with a more accurate spacecraft velocity model indicates that the
    velocity scale for the data presented here has an uncertainty of
    $\leq 0.3$~km\ s$^{-1}$.

\section{Results and discussion}\label{results}

\begin{figure*}[ht]
   \centering
   \includegraphics[angle=-90, width=17cm]{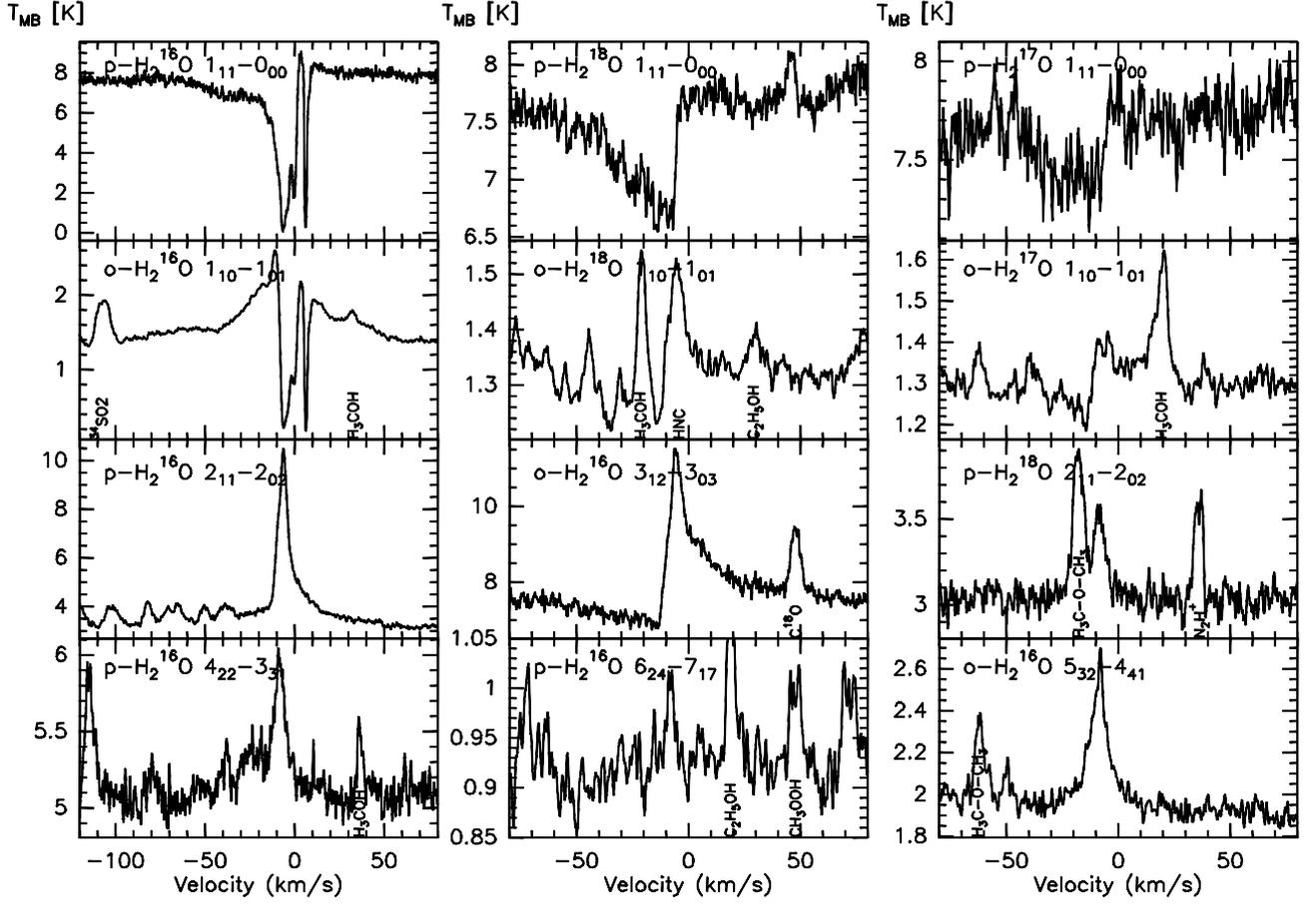}
   \caption{Spectra of twelve water lines in the high-mass star-forming
     region NGC~6334~I. \emph{Top row:} Ground-state para transitions
     of H$_2^{16}$O, H$_2^{18}$O and H$_2^{17}$O. \emph{Second row:}
     Ground-state ortho transitions of H$_2^{16}$O, H$_2^{18}$O and
     H$_2^{17}$O. \emph{Third row:} Water transitions with moderate
     upper state energies. \emph{Bottom row:} Water lines with high
     upper state energies ($>400$~K).}
   \label{SpecI}
 \end{figure*}

The spectra of the twelve observed water lines are displayed in
{\bf{Fig.~\ref{SpecI}}}. The shapes of these lines are in general very
complex, indicating the complex structure of the source itself. They
can be divided into three main components: the hot core, the cold
foreground component, and the outflow. Although the spectrum of
NGC~6334~I is in general very rich in emission lines, the water
spectra presented here are relatively clean. Emission lines of other
molecules interfering with the H$_2$O lines are only seen near
H$_2^{18}$O~$2_{11}-2_{02}$ (H$_3$C-O-CH$_3$) and
H$_2^{18}$O~$1_{10}-1_{01}$ (HNC and H$_3$COH).

Below we discuss the three individual components present in the water spectra.

\subsection{Cold foreground gas}\label{cg}

The foreground material as seen in the ground-state ortho- and
para-H$_2^{16}$O transitions, consists of three velocity components at
$-6.3\pm 0.3$, $-0.3\pm0.12$, and $+6.2\pm0.15$ km\,s$^{-1}$. All
    three components appear in absorption against the strong continuum
    emitted by the warm dust in the hot core of NGC~6334~I. The
parameters of the absorption features seen in the 1$_{11}$-0$_{00}$
and the 1$_{11}$-1$_{01}$ spectra are listed in Table~\ref{lpfor}. The
velocity of the different components are derived from Gaussian fits.
However, Gaussian fits do not correctly match the depth of the absorption. Therefore the other parameters are determined by
    visually inspecting the spectra and integrating over the relevant
    velocity ranges. To derive the continuum level we applied a
linear baseline fit in the vicinity of the absorption line. Thus the
background radiation contains not only the dust continuum, but also
water emission from all the components in the background (hot core and
outflow). The velocity-averaged optical depths of the individual
components derived from the line/continuum ratio of the
$1_{11}-0_{00}$ line (ground-state para transition) are $2.1\pm0.83$,
$1.27\pm0.29$ and $1.39\pm0.63$ for the -6.3~km\,s$^{-1}$,
-0.3~km\,s$^{-1}$, and the +6.2~km\,s$^{-1}$ component, respectively.
The corresponding optical depths of the ground-state ortho transition
are $1.79\pm0.53$, $0.94\pm0.20$, and $1.18\pm0.38$. The velocity
ranges over which the optical depths were averaged are given in
Table~\ref{op}. Deriving the optical depth from the line-to-continuum
ratio is based on the assumption that the excitation temperature is
negligible with respect to the continuum brightness temperature, i.e.,
that the foreground component is cold with respect to the background.
The large uncertainties arise because the optical depth of
deep absorption lines is very sensitive to the actual continuum level,
which is determined with an accuracy of $\sim5\%$ in our HIFI spectra.
In addition, the frequency of the H$_2$O~1$_{11}$-1$_{01}$ (557~GHz)
is close to the edges of the bands 1a and 1b. Thus the uncertainty is
enhanced by variations in the sideband ratio of the receiver. For the present
calculation we used a sideband imbalance of 5\%, derived from the
differences of the spectra taken in bands 1a and 1b. The lack of
absorption of H$_2^{18}$O $1_{11}-0_{00}$ and H$_2^{18}$O
$1_{01}-1_{10}$ at +6.2~km\,s$^{-1}$ gives 1$\sigma$ upper limits
for the optical depth of the para- and ortho-H$_2$O ground-state
transitions of 4.5 and 6.2, respectively, assuming an
H$_2^{16}$O/H$_2^{18}$O ratio of 500, which constrains the optical
depth at the line center even more.

Because almost the complete continuum signal is absorbed at -6.3 and 6.2~km\
s$^{-1}$, the absorbing material must be cold. Upper limits for the
excitation temperature are 5.5~K and 9~K for the o-H$_2$O and
p-H$_2$O, respectively. These upper limits are derived assuming large
optical depths, thus the remaining signal at the center of the water
lines corresponds to the source function of the transition, and
therefore reflects the water excitation temperature.

In the following analysis we assume that all water molecules are in
the ortho- and para-ground states. This assumption is justified given
the very high critical densities of these two transitions
($\rm2\cdot10^8~cm^{-3}$ and $\rm1.5\cdot10^7~cm^{-3}$ for the
ground-state para- and ortho-transitions, respectively). The column
densities per velocity interval are thus given by the formula
$$ 
N=\frac{8\pi\cdot\tau\cdot \nu^3}{ c^3 A}\frac{g_l}{g_u},
$$
where $N$ is the column density, $\tau$ the optical depth, $\nu$ the
line frequency, and c is the speed of light. $A$ stands for the
Einstein A coefficient and $g_l$ and $g_u$ are the degeneracy of the
lower and the upper level of the transition. Subsequently, we
integrated over the velocity ranges given in Table~\ref{op} to
determine the column density for each component. For the
-6.3~km\ s$^{-1}$ component we considered only velocities $>$ -6~km\
s$^{-1}$, because at velocities $<$ -6~km\ s$^{-1}$ the signal may be
contaminated by absorption in the outflow. The derived column
densities and the ortho/para (o/p) ratio of the three components are listed in
Table~\ref{op}. Because the uncertainties of the column densities are
dominated by systematical errors (mainly by the uncertainty of the
continuum level and baseline standing waves), the errors given in
Table~\ref{op} are maximum uncertainties.
\begin{table}[ht]
\caption{Column densities of ortho and para water in the foreground material. }             
\label{op}      
\centering                         
\begin{tabular}{c c c c c}        
\hline\hline               
Comp. &  vel. range &  N(o-H$_2$O) & N(p-H$_2$O) & o/p ratio \\  
& kms$^{-1}$ & $10^{12}$~[cm$^{-2}$] &  $10^{12}$~[cm$^{-2}$]  & \\
\hline                        
   -6.3 & -6 to -2 & $31.4\pm9.4$ & $19.3\pm7.8$  & $1.6\pm0.8$ \\     
   -0.3 & -2 to 1  &  $14.2\pm3.1$ & $8.8\pm2.0$ & $1.6\pm0.5$ \\
   +6.2 &  5 to 8  & $17.8\pm5.8$  & $9.6\pm4.3$  & $1.8\pm1.0$ \\
\hline                                   
\end{tabular}
\end{table}
The o/p ratio seems to be lower than three, the value expected in
statistical equilibrium, in all three foreground components. However,
the uncertainties of the derived ratios are quite large for these
highly saturated lines, and thus the derived o/p
ratios have to be treated with caution.

Water is expected to be formed with an o/p ratio of three.
Subsequently, collisions with atomic and molecular ions (e.g., H$^+$,
H$_3^+$) can lead to proton exchanges, and thus thermalization of the
o/p ratio. In cool, dense gas, like the foreground components, an o/p
ratio lower than three is thus likely. From the o/p ratio of
$1.6\pm0.5$ derived for the -0.3~km\ s$^{-1}$ component, which has the
smallest uncertainty, we calculate a spin temperature of $19\pm5$~K,
indicating that the temperature of the foreground component is $\sim
20$~K. An o/p ratio lower than three has also been measured
toward Sgr~B2(M) (Lis et al.~\cite{LPG10}; see this paper for further
discussion).


\subsection{Hot core}

Spectra of H$_2^{16}$O~$4_{22}-3_{31}$, H$_2^{16}$O~$6_{24}-7_{17}$,
and H$_2^{16}$O~$5_{32}-4_{41}$, the three transitions with upper
state energies above 400 K, show only a single component at a velocity
of about $-8.2$~km\ s$^{-1}$, emitted by the hot core of NGC~6334~I.
The parameters of these lines are listed in Table~\ref{Hclin}.
    Integrated intensities are derived by integrating over a velocity
    range, whereas the other parameters are derived by a Gaussian
    fit. This velocity is about 1.5~km\ s$^{-1}$ lower than the
systematic velocity of NGC~6334~I, but is consistent with the velocity
seen in CH $J=\frac{3}{2}-\frac{1}{2}$ (van der Wiel et
al.~\cite{vvL10}) and HCO$^+$~$J=12-11$. Interferometric observations
of NH$_3$~(3,3) and (6,6) reveal a velocity of -8.1~km\ s$^{-1}$ for
one of the embedded cores (mm~2), which shows the highest optical
depth in highly excited ammonia lines (Beuther et al.~\cite{BWT07}).
In addition to the high-energy H$_2^{16}$O lines the
H$_2^{18}$O~$2_{11}-2_{02}$ line (E$\rm _{up}=136.4~K$) peaks at
-8.6~km\ s$^{-1}$ (Table~\ref{Hclin}), suggesting that this
optically thin line is predominantly emitted from the hot core as
well.

Two of the observed lines, H$_2$O~$5_{32}-4_{41}$ and
H$_2$O~$4_{22}-3_{11}$ are expected to show maser activity over a wide
range of physical conditions (Maercker et al.~\cite{MSO08}; Neufeld \&
Melnick~\cite{NM91}), therefore we excluded these lines from the
subsequent analysis. To calculate the excitation temperature and water
abundance in the hot core, we assumed that the two remaining lines are
optically thin and that the levels are populated according to the 
Boltzmann distribution. Furthermore, we adopted an o/p ratio
of three, an H$_2^{16}$O/H$_2^{18}$O of 500, and, based on
interferometric observations (Hunter et al.~\cite{HBM06}) a source
size of $\sim 10''$ to correct for the difference in the beam size. With
these assumptions, we derive a water excitation temperature (T$\rm
_{ex}$) of $217\pm30$~K, and a total H$_2^{16}$O column density of
$\rm 7.5\pm1.0\cdot 10^{15}~cm^{-2}$.

The C$^{18}$O~$J=10-9$ line is observed very close to the
o-H$_2^{16}$O~$3_{12}-3_{03}$ line ($\Delta\nu = -202$~MHz) and allows
us to independently estimate the H$_2$ column density of the hot core.
The integrated intensity of this C$^{18}$O line is 8.78~Kkms$^{-1}$,
which results, assuming a similar excitation temperatures for
C$^{18}$O and H$_2$O and local thermal equilibrium, in a C$^{18}$O
column density of $\rm 7.2\pm1\cdot 10^{14}~cm^{-2}$. This gives
N(H$_2$)=$\rm 3.8\pm0.5\cdot 10^{21}~cm^{-2}$, adopting a CO abundance
of $9.5\cdot 10^{-5}$ and a C$^{16}$O/C$^{18}$O ratio of 500, leading
to a water abundance of $2.0\pm 0.3\cdot 10^{-6}$. The assumption of a
similar excitation temperature of H$_2$O and C$^{18}$O is questionable
considering the large difference in their dipole moments (C$^{18}$O is
likely tracing a slightly more extended, cooler region). Sandell~(\cite{S00}) derives a dust temperature of 100~K for NGC~6334~I.
Assuming 100~K as lower limit for T$\rm_{ex}$ of C$^{18}$O leads to a
C$^{18}$O column density of $1.6\cdot 10^{15}~cm^{-2}$, and
subsequently to a lower limit of $8\cdot 10^{-7}$ for the water
abundance.

\subsection{Outflow}

\begin{figure*}[ht]
   \centering
   \includegraphics[angle=-90, width=14cm]{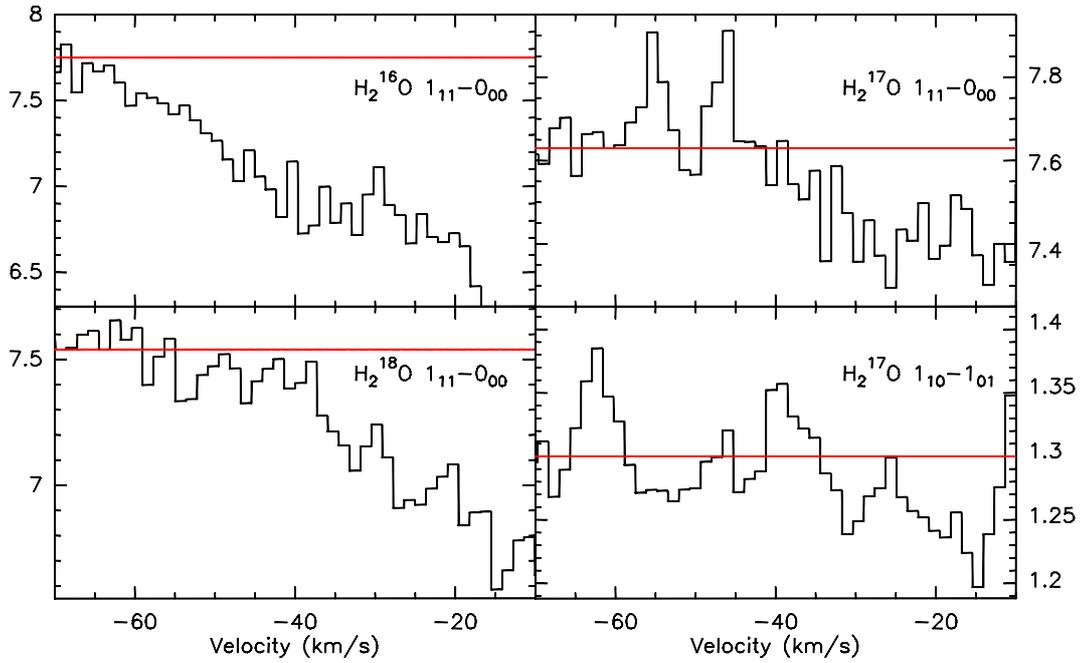}
   \caption{Absorption feature of the blue lobe of the outflow. The
     spectra of the ground-state transitions of p-H$_2^{16}$O,
     p-H$_2^{18}$O, p-H$_2^{17}$O, and o-H$_2^{17}$O are displayed.
     The red line shows the dust continuum level.}
   \label{onlfig}
 \end{figure*}

Outflow features are seen in the ground-state ortho- and para-transitions of all three water isotopologues, except the
H$_2^{18}$O~$1_{10}-1_{01}$ line, which is contaminated by H$_3$COH
and HNC lines. Furthermore, the outflow can be seen in the
H$_2^{16}$O~$2_{11}-2_{02}$ and H$_2^{16}$O~$3_{12}-3_{03}$ lines
(P-Cygni profiles). The spectrum of H$_2^{16}$O~$1_{10}-1_{01}$ is very
interesting because it shows indication of two outflow components, a
broad pedestal ranging from $-90$ to +60~km\ s$^{-1}$ and a narrower,
but brighter component from $-40$ to +20~km\ s$^{-1}$ (see
    Fig.~\ref{onlfig}). Because only one outflow is seen in
interferometric observations (Beuther et al.~\cite{BWT08}) and the
central velocities of the two components are similar, the observed
spectra likely reflect a temperature or density variation within the
outflow.

That the blue outflow lobe shows up in emission in
H$_2^{16}$O~$1_{10}-1_{01}$, but weakly in absorption in
H$_2^{17}$O~$1_{10}-1_{01}$ may be explained by a higher excitation
temperature of the optically thick H$_2^{16}$O line, caused by photon
trapping, which in turn leads to level thermalization.

\begin{figure*}[ht]
   \centering
   \includegraphics[angle=-90, width=17cm]{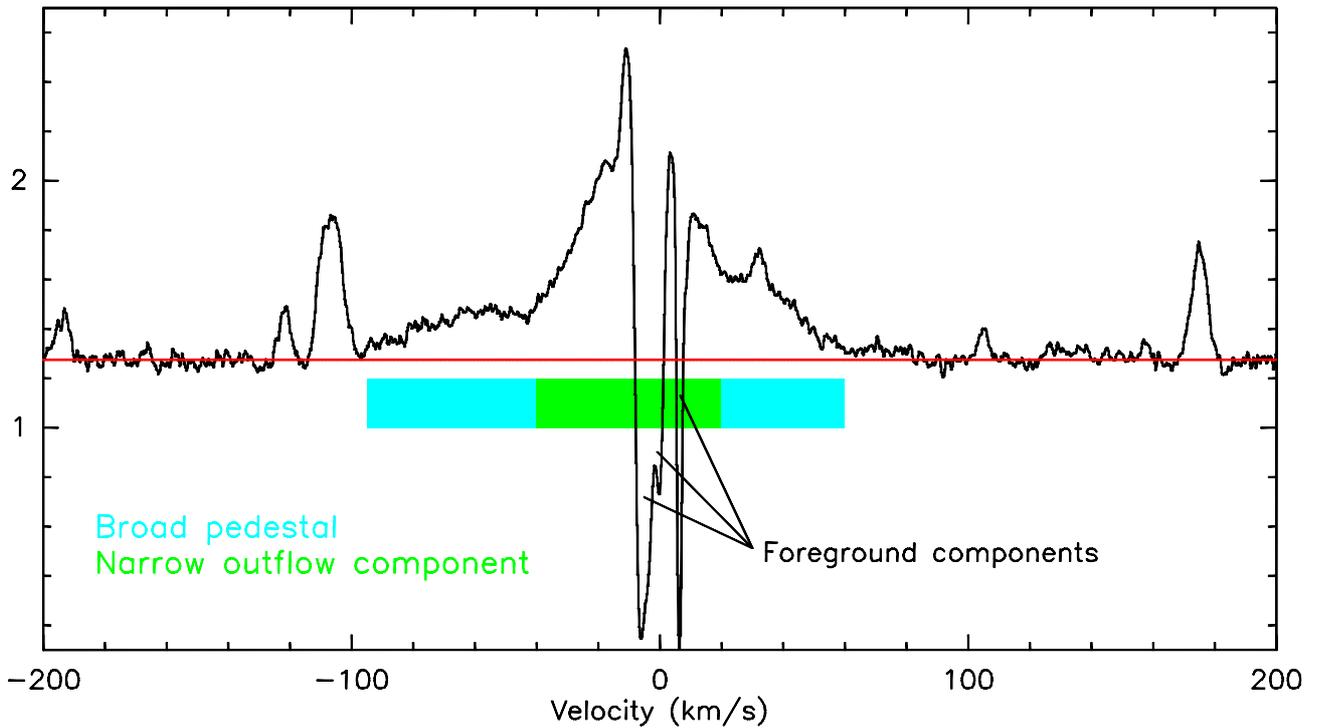}
   \caption{Spectrum of H$_2^{16}$O 1$_{10}$-1$_{01}$. The red line
     marks the dust continuum level. The foreground components and the two outflow components are labeled. The velocity range
     of the broad outflow (pedestal) is marked in light blue and the
     range of the narrow outflow component is marked in yellow.}
   \label{ogf}
 \end{figure*}

A prominent feature is the absorption against the hot core dust
    continuum in the blue lobe of the outflow seen in all
ground-state transitions except H$_2^{16}$O~$1_{10}-1_{01}$ (see
    Fig.~\ref{ogf}). Because absorption features show up even in
H$_2^{17}$O, which is $\approx 1500$ times less abundant than
H$_2^{16}$O, the H$_2^{16}$O lines are likely completely optically
thick. Thus the signal in the H$_2^{16}$O 1$_{11}$-0$_{00}$ line at
velocities from -40~km\ s$^{-1}$ to -20~ km\ s$^{-1}$
(6.82$\pm0.12$~K) is the source function times the beam filling factor
($S\cdot\eta _B$) plus the dust continuum (7.72~K; see
    Fig.~\ref{onlfig}) not affected by absorption. At velocities
higher than -20~ km\ s$^{-1}$ possible absorption by the foreground
component makes the results unreliable, and therefore this velocity
range was excluded from our analysis. From the emission in the red
lobe seen in the H$_2^{16}$O $1_{11}-0_{00}$ transition we conclude
that the $S\cdot\eta _B$ is 0.38~K for H$_2^{16}$O. Therefore
    only a fraction ($f$=17\%) of the total dust continuum is covered
    by the blue lobe of the outflow. This fraction is calculated from
$$
f=1-\frac{T_{MB}-S\cdot\eta _B}{T_c},
$$
where $T_{MB}$ is the detected signal strength and $T_c$ is the level
of the background continuum. No emission is seen from the red lobe
of the 1$_{11}$-0$_{00}$ line of the rarer isotopologues, and therefore
only an upper limit of 0.1~K can be given for the corresponding
$S\cdot\eta _B$. Because this upper limit is of the order of the
measurement uncertainty of the signal of both lines, we calculated the
optical depth and column density of the H$_2^{17}$O $1_{11}-0_{00}$
and H$_2^{18}$O $1_{11}-0_{00}$ lines assuming a source function of
zero (all molecules are in the ortho- or para-ground state). This agrees with the physical parameters found in previous
investigations (e.g., T$\rm _K >15$~K, $\rm n(H_2)>\rm 10^3~cm^{-3}$;
Leurini et al.~\cite{LSP06}). We calculated the optical depth for
    each velocity channel with the formula
$$
\tau=-ln\left(\frac{T_{MB}-(1-f)\cdot T_c}{f\cdot T_c}\right),
$$ 
and the corresponding column density as described in Sect.~\ref{cg}.
The maximum optical depths (v$\rm _{lsr}=-25.7~km\ s^{-1}$) are 0.70
and 0.29 for H$_2^{18}$O and H$_2^{17}$O $1_{11}-0_{00}$,
respectively. The corresponding column densities are $\rm
4.1\pm0.6\cdot 10^{13}~cm^{-2}$ and $\rm1.1\pm0.3\cdot
10^{13}~cm^{-2}$ for p-H$_2^{18}$O and p-H$_2^{17}$O. Hence the
resulting H$_2^{18}$O/H$_2^{17}$O ratio is $3.7\pm0.6$, which agrees well with the values of the $^{18}$O/$^{17}$O ratio reported in
the literature (e.g., Wilson \& Rood~\cite{WR94}). Assuming the same
continuum coverage for the H$_2^{17}$O~$1_{10}-1_{01}$ line, we
calculated the o-H$_2^{17}$O column density in a similar way,  which leads
to an optical depth of 0.34, which in turn resulted in a column density of
$\rm2.7\pm1\cdot 10^{13}~cm^{-2}$. Thus the o/p ratio in the
outflow is $2.5\pm0.8$. The higher o/p ratio in the outflow is
consistent with the gas being warmer ($>50$~K) compared to 
the foreground clouds. Comparing these values to the total column
density of the outflow determined by Leurini et al.~(\cite{LSP06}),
and assuming an H$_2^{16}$O/H$_2^{18}$O ratio of 500, we derive a
water abundance of $4.3\cdot 10^{-5}$, a value typical for warm
($>100~K$) gas.

\section{Summary}\label{con}

\emph{Herschel}/HIFI studies of multiple lines of water isotopologues give
important insights into the physical processes in dense
molecular material. Our observations of NGC~6334~I indicate that

\begin{itemize}

\item H$_2$O lines show complex line profiles, with multiple emission and
  absorption features, originating from multiple spatial components. 

\item the H$_2$O abundance varies between about $4\cdot 10^{-5}$ in the
  outflow, $2\cdot10^{-6}$ in the hot core, and $10^{-8}$ in the cool
  foreground gas. 
  These abundances are in the range of previously determined ISM
  values.

\item the H$_2^{18}$O/H$_2^{16}$O and H$_2^{17}$O/H$_2^{16}$O ratios are
  comparable to the $^{18}$O/$^{16}$O and $^{17}$O/$^{16}$O isotopic
  ratios determined from observations of the CO isotopologues.

\item the water o/p ratio appears to be lower than the
  statistical value of three in the cold, quiescent material, whereas
  it is close to three in the warmer outflow gas. The lower
  o/p ratio in the foreground component may be explained by proton
  exchange reactions with H$^+$ and H$_3^+$, which lower the o/p ratio
  to a value corresponding to the gas kinetic temperature of about 20~K.

\end{itemize}

Observations of many additional water lines will be available in the
future, when the complete spectral scan of NGC~6334~I is carried out.
This will allow a more detailed modeling of water abundances in this
interesting source.


\hspace{-0.5cm}\rule{9cm}{0.3mm}

\begin{acknowledgements}
  HIFI has been designed and built by a consortium of institutes and
  university departments from across Europe, Canada and the United
  States under the leadership of SRON Netherlands Institute for Space
  Research, Groningen, The Netherlands and with major contributions
  from Germany, France abd the US. Consortium members are: Canada:
  CSA, U.Waterloo; France: CESR, LAB, LERMA, IRAM; Germany: KOSMA,
  MPIfR, MPS; Ireland, NUI Maynooth; Italy: ASI, IFSI-INAF,
  Osservatorio Astrofisico di Arcetri-INAF; Netherlands: SRON, TUD;
  Poland: CAMK, CBK; Spain: Observatorio Astron\'omico Nacional (IGN),
  Centro de Astrobiolog\'{\i}a (CSIC-INTA). Sweden: Chalmers
  University of Technology - MC2, RSS \& GARD; Onsala Space
  Observatory; Swedish National Space Board, Stockholm University -
  Stockholm Observatory; Switzerland: ETH Zurich, FHNW; USA: Caltech,
  JPL, NHSC.  We thank many funding agencies for financial support.
\end{acknowledgements}
\phantom{slksflgh}
$^1$
California Institute of Technology, Pasadena, USA\\
$^2$
Universit\'{e} de Bordeaux, Laboratoire d’Astrophysique de Bordeaux, France; CNRS/INSU, UMR 5804, Floirac, France\\
$^3$
Laboratoire d’Astrophysique de Grenoble, UMR 5571-CNRS, Universit\'e Joseph Fourier, Grenoble France\\
$^4$
INAF - Istituto di Fisica dello Spazio Interplanetario, Roma, Italy\\
$^5$
Infared Processing and Analysis Center,  Caltech, Pasadena, USA\\
$^6$
CESR, Universit\'e Toulouse 3 and CNRS, Toulouse, France\\
$^7$
School of Physics and Astronomy, University of Leeds, Leeds UK\\
$^8$
Observatoire de Paris-Meudon, LERMA UMR CNRS 8112. Meudon, France\\
$^9$
Centro de Astrobiolog\`{\i}a, CSIC-INTA, Madrid, Spain\\
$^{10}$
Max-Planck-Institut f\"{u}r Radioastronomie, Bonn, Germany\\
$^{11}$
INAF Osservatorio Astrofisico di Arcetri, Florence Italy\\
$^{12}$
Astronomical Institute 'Anton Pannekoek', University of Amsterdam, Amsterdam, The Netherlands\\
$^{13}$
Department of Astrophysics/IMAPP, Radboud University Nijmegen,  Nijmegen, The Netherlands\\
$^{14}$
IGN Observatorio Astron\'{o}mico Nacional, Alcal\'{a} de Henares, Spain\\
$^{15}$
Jet Propulsion Laboratory,  Caltech, Pasadena, CA 91109, USA\\
$^{16}$
SRON, Groningen, The Netherlands\\
$^{17}$
Ohio State University, Columbus, OH, USA\\
$^{18}$
Center for Astrophysics, Cambridge MA, USA\\
$^{19}$
Johns Hopkins University, Baltimore MD,  USA\\
$^{20}$
Physikalisches Institut, Universit\"{a}t zu K\"{o}ln, K\"{o}ln, Germany\\
$^{21}$
Institut de RadioAstronomie Millim\'etrique, Grenoble - France\\
$^{22}$
Leiden Observatory, Leiden University, Leiden, The Netherlands\\
$^{23}$
Department of Physics and Astronomy, University College London, London, UK\\
$^{24}$
INAF - Osservatorio Astronomico di Roma, Monte Porzio Catone, Italy
$^{25}$
Department of Astronomy, University of Michigan, Ann Arbor, USA
$^{26}$
Max-Planck-Institut für Astronomie, Heidelberg, Germany
$^{27}$
Max-Planck-Institut für Sonnenphysikforschung, Lindau, Germany
\clearpage

\clearpage

\begin{table*}[ht]
\caption{Summary of the observed lines}             
\label{lines}      
\centering                         
\begin{tabular}{c c c c c}        
\hline\hline               
Transition & spin & E$\rm _{low}$ [K] & $\nu$ [GHz] & beam size [$''$] \\
\hline                        
H$_2^{16}$O~$6_{24}-7_{17}$ & p & 844.2 & 488.491 & 44\\
H$_2^{18}$O~$1_{10}-1_{01}$ & o & 34.3 & 547.676 & 41\\
H$_2^{17}$O~$1_{10}-1_{01}$ & o & 34.3 & 552.021 & 41\\
H$_2^{16}$O~$1_{10}-1_{01}$ & o & 34.3 & 556.936 & 41\\
H$_2^{16}$O~$5_{32}-4_{41}$ & o & 702.0 & 620.701 & 33\\
H$_2^{18}$O~$2_{11}-2_{02}$ & p & 100.7 & 745.320 & 30\\
H$_2^{16}$O~$2_{11}-2_{02}$ & p & 100.9 & 752.033 & 30\\
H$_2^{16}$O~$4_{22}-3_{31}$ & p & 410.6 & 916.172 & 25\\
H$_2^{16}$O~$3_{12}-3_{03}$ & o & 196.9 & 1097.365 & 20\\
H$_2^{18}$O~$1_{11}-0_{00}$ & p & 0.0 & 1101.698 & 20\\
H$_2^{17}$O~$1_{11}-0_{00}$ & p & 0.0 & 1107.167 & 20\\
H$_2^{16}$O~$1_{11}-0_{00}$ & p & 0.0 & 1113.343 & 20\\
\hline                                   
\end{tabular}
\end{table*}

\vspace{2cm}

\begin{table*}[ht]
  \caption{Parameters of the narrow absorption features seen in the
    two ground-state H$_2^{16}$O transitions. $\int T_{MB} dv$ is the
    intensity integrated over the velocity ranges of the individual
    components (Table~\ref{op}). The continuum level given  is an
    average value for the individual components.}             
\label{lpfor}      
\centering                         
\begin{tabular}{c c c c c}        
\hline\hline               
Line & v$\rm _{lsr}$ & T$\rm _{MB}$ & $\int T_{MB} dv$ & cont. level \\
     & km/s & K & Kkm\ s$^{-1}$ & K\\
\hline                        
 & -6.2$\pm 0.3$ & 0.21$\pm0.17$ & 6.2$\pm0.31$ & 7.7$\pm0.4$\\
1$_{11}$-0$_{00}$  & -0.34$\pm 0.1$ & 1.87$\pm0.17$  & 7.8$\pm0.39$ & 7.7$\pm0.4$\\
 & +6.1$\pm 0.10$ & 0.33$\pm0.17$ &  7.8$\pm0.39$ & 9.0$\pm0.45$\\
\hline
 &  -6.8$\pm 1.0$ & 0.22$\pm0.024$ &  2.04$\pm0.10$ & $1.54\pm0.08$\\
1$_{10}$-1$_{01}$  & -0.25$\pm 0.1$ & 0.81$\pm0.024$ & 2.69$\pm0.13$ & $1.53\pm0.08$ \\
 &  +6.3$\pm 0.10$ & 0.19$\pm0.024$ & 2.36$\pm0.76$ & $1.51\pm0.08$\\
\hline
\hline                                   
\end{tabular}
\end{table*}

\vspace{2cm}

\begin{table*}[ht]
\caption{Hot core line parameters}             
\label{Hclin}      
\centering                         
\begin{tabular}{c c c c c}        
\hline\hline               
Line &  $\int T_{MB} dv$ & v$_{lsr}$ & $\Delta$v & T$_{MB}$  \\   
     &    K~kms$^{-1}$ & kms$^{-1}$ & kms$^{-1}$ & K \\
\hline                        
   H$_2^{16}$O 4$_{22}$-3$_{31}$  & 4.96$\pm0.11$  & -8.3$\pm0.11$ & 6.1$\pm0.27$ & 0.66$\pm0.04$\\     
   H$_2^{16}$O 6$_{24}$-7$_{17}$  & 0.28$\pm0.04$ & -8.1$\pm0.16$ & 3.8$\pm0.41$ & 0.084$\pm0.012$\\
   H$_2^{16}$O 5$_{32}$-4$_{41}$  & 4.80$\pm0.06$ & -8.6$\pm0.60$ & 7.8$\pm0.14$ & 0.522$\pm0.013$\\
   H$_2^{18}$O $2_{11}-2_{02}$ & 3.2$\pm0.2$ & -8.6$\pm0.3$ &  6.3$\pm0.5$ & 0.46$\pm0.05$ \\
\hline                                   
\end{tabular}
\end{table*}

\end{document}